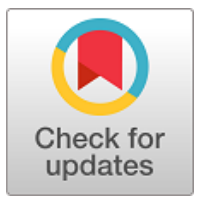

# Loss-robust crossband entanglement generation beyond the direct-transduction limit

**HAOWEI SHI**[1,3] AND **QUNTAO ZHUANG**[1,2,4]

[1] *Ming Hsieh Department of Electrical and Computer Engineering, University of Southern California, Los Angeles, California 90089, USA*
[2] *Department of Physics and Astronomy, University of Southern California, Los Angeles, California 90089, USA*
[3] *haow.shi@gmail.com*
[4] *qzhuang@usc.edu*

**Entanglement across distant frequency bands is a crucial resource in quantum networking. However, directly entangling crossband photons, e.g., microwave and optical, is challenging. Furthermore, distributing crossband entanglement via direct quantum transduction is fundamentally limited, regardless of input engineering with unconstrained source brightness. We propose to utilize intraband entanglement to overcome such direct-transduction limits by a factor that increases with the input intraband entanglement brightness in the ideal case. In the presence of experimental loss 5% and assuming 10 dB of squeezing in both optical and microwave bands, we show that our protocol can generate a violation of the separability criterion equivalent to 2.38 ebits, compared with the baseline protocol limited to 0.082 ebits. The proposed protocols rely only on off-the-shelf components and provide advantages robust to a substantial amount of loss.**

## 1. INTRODUCTION

Entanglement is an essential resource in quantum science and engineering, enabling applications such as quantum-enhanced sensing, quantum communication, and quantum processors in computing [1–5]. Owing to the relatively low loss and low noise of light propagation, optical entanglement can be generated and detected at a distance of thousands of kilometers, which is crucial in long-distance quantum communication [6] and testing of fundamental physics [7]. In these scenarios, entanglement is intraband—the frequencies of the entangled photons are either identical (time-bin entanglement) or close (spontaneous-parametric down-conversion).

While intraband entanglement already provides numerous benefits, crossband entanglement between different frequency bands is more versatile for quantum networking [8–11]. The fundamental reason is that photons at different frequencies behave quite differently under current technology. Microwave photons interact strongly within superconducting cavities to enable quantum computation, and yet microwave photons face huge challenges in transmission, due to the abundant room-temperature noise photons at microwave frequency and the high loss of coaxial cables. Therefore, to entangle distant quantum computers operating at microwave frequencies for distributed quantum computing, crossband entanglement between optical and microwave is crucial. By transmitting two flying optical photons to a relay node for joint detection, an entanglement-swapping process can convert two pairs of microwave-optical entangled photons to microwave–microwave entanglement [12]. Similarly, generating optical–microwave entanglement allows versatile quantum communication between different frequencies via teleportation [13–16].

However, entangling crossband photons across orders-of-magnitude different frequencies, e.g., between microwave and light, is extremely challenging, due to the weak nonlinearity of photons [17]. Alternatively, one can distribute entanglement through direct quantum transduction [18–33]. However, the efficiency and bandwidth of direct quantum transduction are far from satisfactory [34]. Furthermore, a fundamental limit [35] indicates that quantum communication rate of direct quantum transduction—including the crossband entanglement generation rate—is constrained by the channel transmissivity (transduction efficiency) regardless of input engineering and source brightness, due to the quantum information leakage to the environment in the lossy channel.

In this work, we resolve this problem by engineering the environment, and propose to enhance the generation of crossband entanglement by injecting intraband entanglement. A key insight is that quantum transduction goes beyond point-to-point channel communication, as the environment mode, e.g., the optical input in a microwave-to-optical transduction, is in fact controllable. In addition, transduction happens in a local set-up, the input and output are in fact within a single device, allowing direct cooperation. Different from previous encoding-specific approaches relying on complex non-Gaussian environment [36], adopting the sandwiched intraband two-mode squeezing structure proposed for enhancing direct transduction [37], we propose an entirely Gaussian approach capable of generating Gaussian entanglement that is directly applicable in near-term applications [38–40], which achieves an enhanced entanglement generation rate proportional to the input intraband entanglement squeezing levels (in dB) asymptotically. The advantage over the direct-transduction limit survives even in the presence of high coupling loss.

## 2. ENTANGLEMENT MEASURES

Now we briefly introduce the entanglement measures to be used in this paper, while leaving the full details in Appendix A. For a bipartite pure state $|\psi\rangle_{AB}$, entanglement is directly quantified by the entanglement entropy [41], the von Neumann entropy of either subsystem $S(A) = S(B)$, which coincides with the distillable entanglement [41] and entanglement of formation [42]. However, for mixed states, an efficiently calculable entanglement measure is elusive, and we adopt the logarithmic negativity [43], which quantifies the maximum violation of the separability criterion for two-mode Gaussian states [44,45] and is an upper bound on distillable entanglement.

An alternative entanglement measure is the Einstein–Podolsky–Rosen (EPR) quadrature squeezing, which is experiment-friendly [17] and especially useful in quantum sensing [38–40]. We explain EPR quadrature squeezing via the two-mode squeezed vacuum (TMSV) state, a common example of the continuous-variable entanglement. Given two modes $\hat{a}_1, \hat{a}_2$ in TMSV, the variances of its balanced EPR quadratures, $\hat{q}_- \equiv \mathrm{Re}(\hat{a}_1 - \hat{a}_2)$ and $\hat{p}_+ \equiv \mathrm{Im}(\hat{a}_1 + \hat{a}_2)$, are squeezed below the vacuum fluctuation. For such EPR-type entanglement, a natural entanglement measure is the average variance of the EPR quadratures, defined as $\Delta^-_{\mathrm{EPR}} \equiv (\mathrm{var}\ \hat{q}_- + \mathrm{var}\ \hat{p}_+)/2$. In the presence of loss, the balanced EPR quadrature is no longer a good benchmark; instead, the minimal EPR quadrature variance product, where the EPR quadratures are defined with an optimized unbalanced beamsplitter, can be given by the logarithmic negativity [45].

## 3. OVERVIEW

To benchmark existing entanglement generation protocols, we take electro-optic system as an example and define the baseline protocols without intraband entanglement assistance, including direct-transduction distribution with red-detuned pumping and direct entanglement generation with blue-detuned pumping. While the direct-transduction rate is limited by intrinsic coupling efficiency $\eta$, due to the Pirandola–Laurenza–Ottaviani–Banchi (PLOB) bound [35] proportional to $\eta$, we show that direct entanglement generation is subject to a similar limit. Importantly, the intrinsic coupling efficiency $\eta$ is fully determined by the nonlinear coupling coefficient and the pumping strength; the former is limited by the material, and the latter induces noises [46,47]. These limits remain finite even when allowing infinite source brightness. In this paper, we propose to utilize intraband entanglement to overcome such limits, with advantage scaling up with the input intraband squeezing strength and a better rate scaling with $\sqrt{\eta}$ for weak coupling $\eta \ll 1$.

Our first insight is that the entanglement-assisted (EA) transduction protocol in Ref. [37] not only enhances transduction efficiency but also allows simultaneous crossband entanglement.

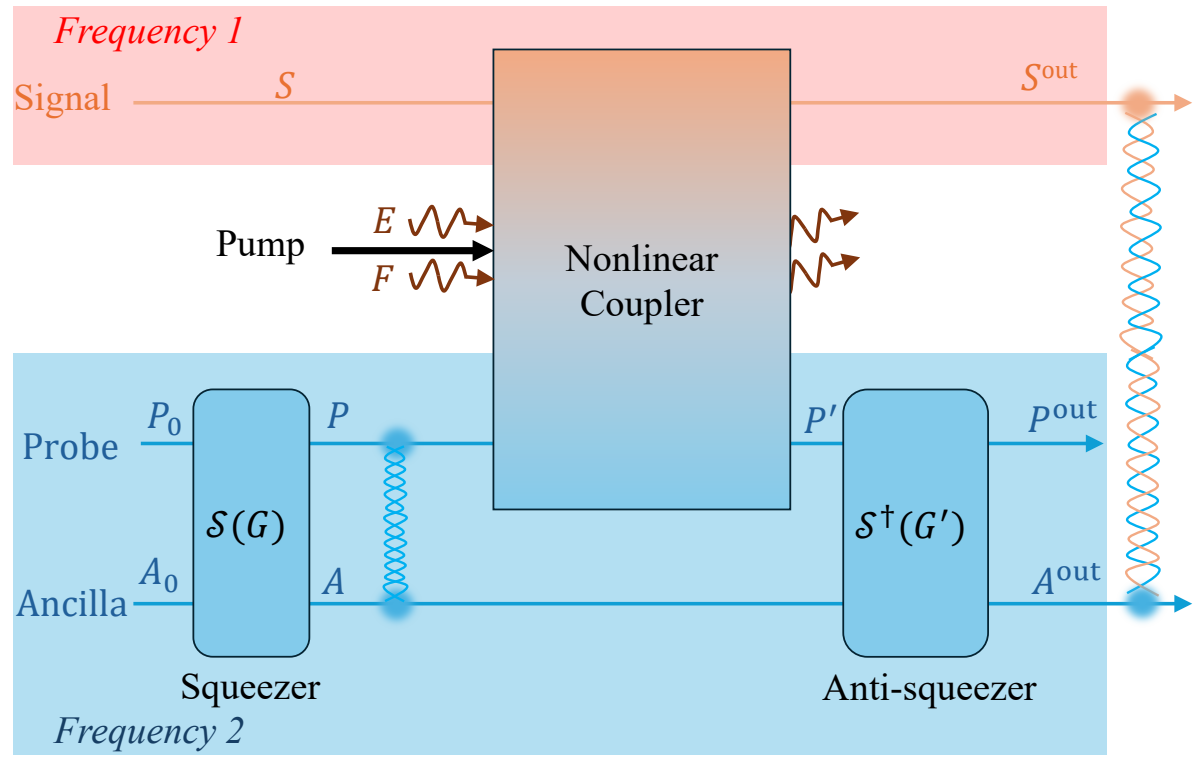

**Fig. 1.** Schematic of simultaneous entanglement generation and transduction using the single-band entanglement-assisted (EA) transduction protocol for entanglement generation and transduction. The two-mode squeezer and anti-squeezer amplify the transduction from $S$ to $P^{\mathrm{out}}$ in the middle noiselessly. As a byproduct, a crossband entanglement is generated between $S^{\mathrm{out}}$ and $A^{\mathrm{out}}$.

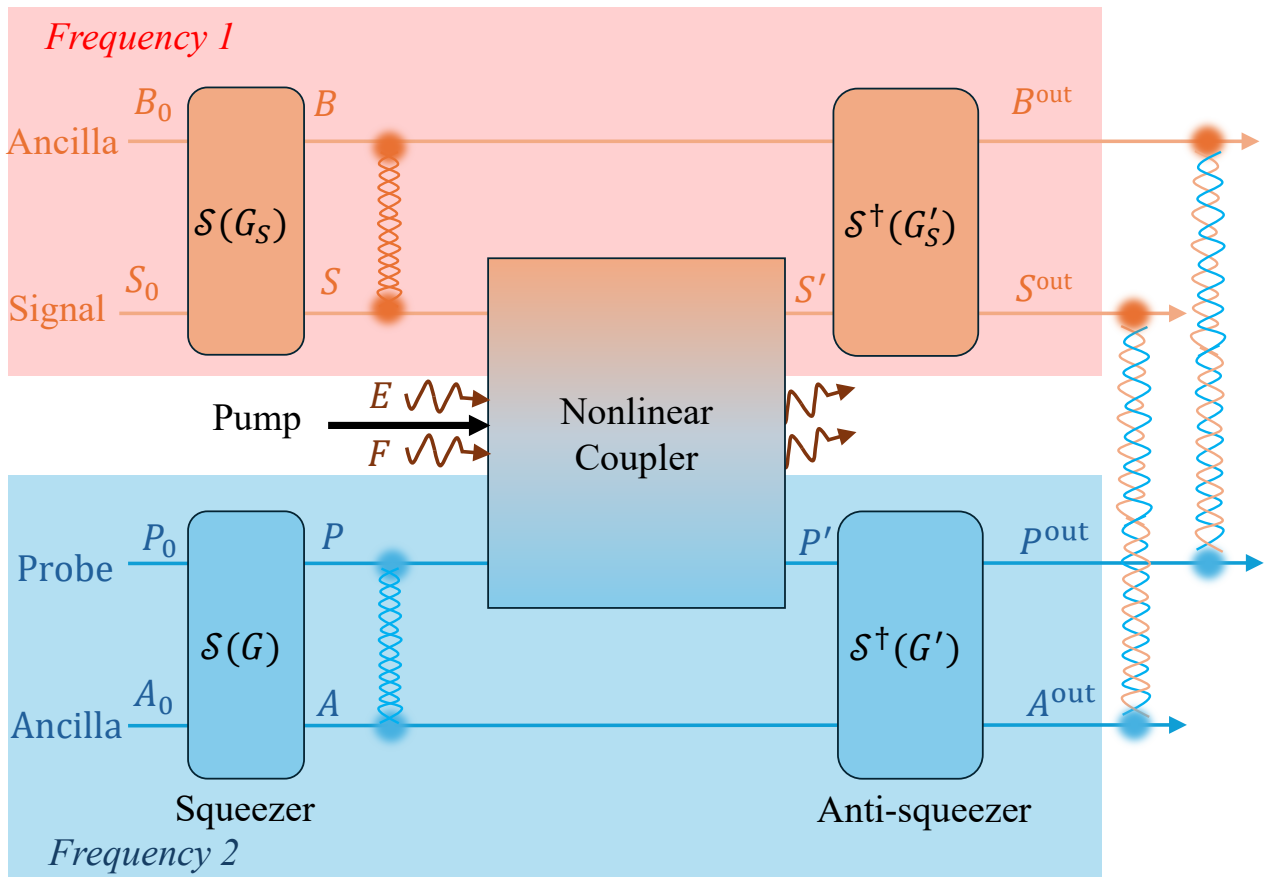

**Fig. 2.** Schematic of dual-band cooperative EA entanglement generation protocol. The setup is similar to Fig. 1, with the signal $S$ at frequency 1 band further entangled with another ancilla $B$, and a matched anti-squeezer $\mathcal{S}^\dagger(G'_S)$ at the output of frequency 1.

As shown in Fig. 1, Ref. [37] proposed the squeezer–coupler–anti-squeezer sandwich structure to noiselessly amplify the coupling in the middle, and thereby amplify the transduction from signal $S$ at frequency band 1 to output probe $P^{\mathrm{out}}$ at frequency band 2. In this paper, we show that crossband entanglement is generated between the $S^{\mathrm{out}}$ and $A^{\mathrm{out}}$ as a beneficial byproduct. Given vacuum signal input $S$ and lossless coupler, the output $S^{\mathrm{out}} A^{\mathrm{out}}$ is in a pure TMSV state with per-mode mean photon number $N_S = \eta(G-1)$ scaling up with input squeezing gain $G$.

To further enhance entanglement generation, we propose a dual-band cooperative EA entanglement generation protocol as shown in Fig. 2, where we introduce ancilla $B_0$ entangled with the signal $S_0$ to utilize intraband entanglement assistance at both frequency bands. In the coupler lossless regime (e.g., an overcoupled cavity), we show that pure crossband entanglement in the form of TMSV states between both pairs ($S^{\mathrm{out}}$, $A^{\mathrm{out}}$) and ($P^{\mathrm{out}}$, $B^{\mathrm{out}}$) can be generated with $N_S \simeq \sqrt{\eta} G$, given symmetric input squeezing gain $G_S = G$, with an improvement by a factor of $1/\sqrt{\eta}$ in the mean photon number compared with the single-band protocol. Such improvement is significant in the weak nonlinear coupling case ($\eta \ll 1$), which is typical in the state-of-the-art microwave-optical transduction and entanglement experiments [17,24,32].

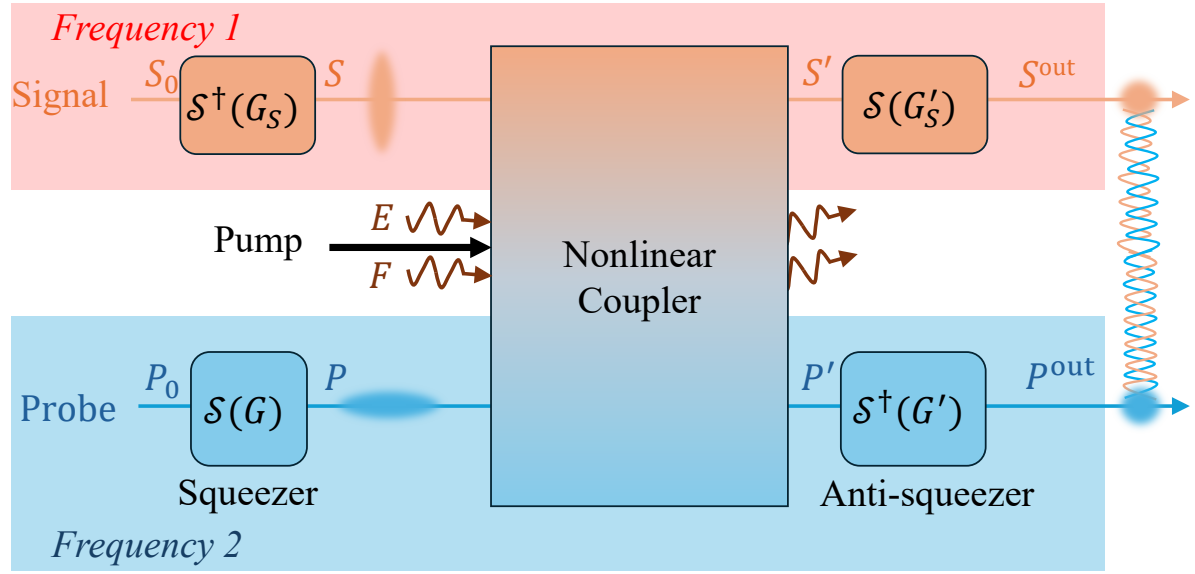

**Fig. 3.** Schematic of ancilla-free entanglement generation protocol based on dual-band cooperative transduction. Compared with Fig. 2, here the ancillae $A$,$B$ are removed, the squeezers and anti-squeezers are single-mode.

To distinguish the EA advantage from the environment control advantage, we also analyze an ancilla-free protocol based on the dual-band cooperative setup, as shown in Fig. 3, where we remove the ancillae and replace the two-mode squeezer/anti-squeezer with the single-mode version.

It is noteworthy that the anti-squeezers in all the protocols are optional if the goal is to generate entanglement between the two frequency bands, without requiring it to be in pairs. This is because these anti-squeezers are local operations within a single frequency band that do not change the crossband entanglement.

In the ideal lossless limit, in Section 9, we verify that our single-band and dual-band EA proposals are able to generate crossband entanglement scaling up with the input squeezing strength $G$. In contrast, the baseline is limited by a constant. A 3-dB intraband squeezing is sufficient to beat the baseline benchmark. Remarkably, the scaling with $G$ of entanglement generation of the ancilla-free protocol is similar to the EA protocols. In the presence of coupling loss, as shown in Section 10, all three proposed protocols yield loss-robust advantages beyond the baseline. The single-band EA protocol and the dual-band ancilla-free protocol yield similar advantages, while the dual-band EA protocol further improves the advantage as a result of combining the EA advantage and the environment control advantage.

## 4. TRANSDUCTION AND ENTANGLEMENT ASSISTANCE

Here, we establish the theory model of quantum transduction, following Ref. [37]. While our results apply to all bosonic transduction platforms including both photon and phonon interactions that generate entanglement, here we take the electro-optic nonlinear coupling as an example, which sets the state of the art in microwave-optical entanglement generation [17]. The interaction Hamiltonian of cavity electro-optics has the standard three-wave mixing form $H = \hbar g \hat{a}^\dagger \hat{a}(\hat{b}^\dagger + \hat{b})$, with an optical mode ($\hat{a}$) and a microwave mode ($\hat{b}$). Here $g$ is the coupling coefficient and $\hbar$ is Planck's constant. In the case of microwave-to-optical transduction, external microwave signal field $\hat{\mathcal{E}}_S$ couples into cavity microwave mode $\hat{b}$, then couples with cavity optical mode $\hat{a}$ via nonlinearity, finally couples out with external optical probe field $\hat{\mathcal{E}}_{P'}$. To fulfill direct entanglement (DE), Ref. [17] pumps the blue sideband of the optical mode $\hat{a}$ with a strong coherent laser to induce a two-mode squeezing interaction between the optical mode $\hat{a}$ and the microwave mode $\hat{b}$. Alternatively, one can pump the optical mode $\hat{a}$ at the red sideband, then a beamsplitter interaction can be realized between $\hat{a}$ and $\hat{b}$ for direct transduction (DT).

In this paper, we will focus on DT. As sketched in Fig. 1, a quantum transduction device is such a beamsplitter-type nonlinear coupler that converts the signal in frequency 1 to the probe at frequency 2. It can be described by the input–output relation [37]:

$$\hat{\mathcal{E}}_{P'} = \sqrt{\kappa}\hat{\mathcal{E}}_P + \sqrt{\eta}\hat{\mathcal{E}}_S + \sqrt{\kappa_E}\hat{\mathcal{E}}_E, \tag{1}$$

where $\kappa$ is the reflectivity from $P$ to $P'$:

$$\eta = \zeta_{\text{m}}\zeta_{\text{o}}\frac{4C}{(1+C)^2}, \tag{2}$$

is the intrinsic transduction efficiency [23] from $S$ to $P'$, where the cooperativity $C \propto g^2$ describes the interaction strength, $\zeta_{\text{m}}, \zeta_{\text{o}}$ are the microwave and optical extraction efficiencies; $\kappa_E = 1 - \eta - \kappa$ is the intrinsic loss induced by $1-\zeta_m$ and $1-\zeta_o$. We assume the environment $\hat{\mathcal{E}}_E$ is cooled to vacuum. The full two-output beamsplitter model for the dual-band protocol is given in Appendix B. The beamsplitter model holds for general bosonic transduction, not necessarily associated to any specific frequency.

In the EA direct-transduction protocol [37], we first prepare the probe and the ancilla using a two-mode squeezer $\mathcal{S}(G)$ of gain $G$ on initial vacuums $P_0$ and $A_0$, which yields the entangled $P, A$ modes with field operator $\hat{\mathcal{E}}_P = \sqrt{G}\hat{\mathcal{E}}_{P_0} + \sqrt{G-1}\hat{\mathcal{E}}^\dagger_{A_0}$, $\hat{\mathcal{E}}_A = \sqrt{G}\hat{\mathcal{E}}_{A_0} + \sqrt{G-1}\hat{\mathcal{E}}^\dagger_{P_0}$. After the two-mode squeezing, the probe goes through Eq. (1) and the ancilla is intact. Finally, the probe and the ancilla are anti-squeezed using $\mathcal{S}^\dagger(G')$: $\hat{\mathcal{E}}_{P^{\text{out}}} = \sqrt{G'}\hat{\mathcal{E}}_{P'} - \sqrt{G'-1}\hat{\mathcal{E}}^\dagger_A$ $\hat{\mathcal{E}}_{A^{\text{out}}} = \sqrt{G'}\hat{\mathcal{E}}_A - \sqrt{G'-1}\hat{\mathcal{E}}^\dagger_{P'}$. To minimize the quantum amplification noise from anti-squeezing, the optimal $G'$ is chosen as $G' \leftarrow G'^\star \equiv 1/(1-\kappa+\kappa/G)$. In this case, the noise background in the output probe is in vacuum state $\hat{\mathcal{E}}^\star_{P^{\text{out}}} = \sqrt{\eta_{\text{EA}}}\hat{\mathcal{E}}_S + \sqrt{1-\eta_{\text{EA}}}\hat{\mathcal{E}}_{\text{VAC}}$, where the background $\hat{\mathcal{E}}_{\text{VAC}}$ is in vacuum state, and:

$$\eta_{\text{EA}} = \eta G'^\star = \eta\frac{G}{G(1-\kappa)+\kappa}, \tag{3}$$

is the noiseless EA transduction efficiency, enhanced by a factor of $G'^\star$. At the limit of $G \to \infty, \kappa_E \to 0$, we have $\eta_{\text{EA}} \to 1$. Here the no-cloning theorem is not violated because the reflected signal system $S^{\text{out}}$ carries the partially transduced TMSV from $P$ and is extremely noisy. In fact, in Section 6, we reveal that $S^{\text{out}}A^{\text{out}}$ are in TMSV state, producing the desired crossband entanglement.

## 5. BASELINE

For direct transduction (DT), direct coherent conversion between the microwave and optical modes is realized by Eq. (1), which forms a bosonic pure loss channel from $S$ to $P'$, with transmissivity $\eta$. To generate entanglement, various protocols can be considered. On the input side, one may prepare arbitrary entanglement between two modes and send one mode through the pure loss channel. On the output side, one may perform arbitrary quantum operations to distill the entanglement. On both input and output sides (which correspond to different frequency bands but are at the same location), we can coordinate the quantum operations with classical communication. Despite all possible DT protocols, the entanglement generation rate is limited by the PLOB bound [35]:

$$Q^{\mathrm{DT}}(\eta) = \log_2\left(\frac{1}{1-\eta}\right). \quad (4)$$

Note that here PLOB bound allows two-way classical communication between operations across the two frequency bands; therefore, using Eq. (4) as the DT limit is more general than the one-way quantum capacity limit considered in Ref. [16].

For direct entanglement (DE), the optical mode and the microwave mode are directly entangled in a noisy TMSV state $\hat{\rho}_{\mathrm{m,o}}$, with zero mean and the DE covariance matrix [16]:

$$V^{\mathrm{DE}}_{\mathrm{m,o}} = \begin{pmatrix} u\boldsymbol{I}_2 & v\boldsymbol{Z}_2 \\ v\boldsymbol{Z}_2 & w\boldsymbol{I}_2 \end{pmatrix}, \quad (5)$$

where $u = 1 + \frac{8\zeta_{\mathrm{m}}C}{(1-C)^2}$, $v = \frac{4\sqrt{\zeta_{\mathrm{o}}\zeta_{\mathrm{m}}C}(1+C)}{(1-C)^2}$, $w = 1 + \frac{8C\zeta_{\mathrm{o}}}{(1-C)^2}$. We have ignored both the optical and microwave thermal noise for simplicity.

To provide a baseline benchmark, i.e., the best performance of the baseline protocols, below we compare DT and DE in terms of the output entanglement entropy. We consider the ideal cavity-overcoupled limit, such that the transduction is lossless $\kappa_E \to 0$. At this limit, it is easy to verify from Eq. (5) that DE generates a pure TMSV state with mean photon number $N_S = 4C/(1-C)^2$; while under the same value of $C$, DT provides a bosonic pure loss channel with conversion efficiency $\eta = 4C/(1+C^2)$, as defined in Eq. (1). Choosing $\eta$ to characterize the cavity, the DE mean photon number $N_S = \frac{\eta}{1-\eta}$. At the overcoupled limit, we find that DE is strictly better than DT, by comparing the entanglement entropy for DE:

$$E^{\mathrm{DE}}_{\kappa_E=0}(\eta) = \log_2\left(\frac{1}{1-\eta}\right) + \frac{\eta}{1-\eta}\log_2\left(\frac{1}{\eta}\right) \geq Q^{\mathrm{DT}}(\eta), \quad (6)$$

with the PLOB bound in Eq. (4) for DT. Therefore, we will take Eq. (6) as the baseline benchmark for the proposed protocols.

Here we observe that DE can already enable overcoming the direct-transduction limit. However, the advantage is quite limited in the range of $\eta > 10^{-2}$, which is the typical range of interest for quantum transduction devices. Additionally, the entanglement generation rate in Eq. (6) is fixed for a given $\eta$, while the proposed protocol in the following sections overcomes the direct-transduction limit with an advantage growing with the squeezing levels.

## 6. SINGLE-BAND ENTANGLEMENT-ASSISTED PROTOCOL

In this section, we show that during the EA transduction process, crossband entanglement simultaneously arises between output signal $S^{\mathrm{out}}$ and ancilla $A^{\mathrm{out}}$, besides the enhanced transduction from input signal $S$ towards the output probe $P^{\mathrm{out}}$, as depicted in Fig. 1. Here we present a simplified analysis assuming no additional loss ($\kappa_E = 0$, $\kappa = 1-\eta$), while the effect of loss will be addressed at the end of the manuscript. An analysis including loss for the symmetric case is in Section 10, and a full analysis is in Appendix C.

Consider the input signal in an arbitrary pure state $|\phi\rangle_S$ with finite energy, the probe and ancilla are in vacuum state. After the first two-mode squeezing between $A_0$ and $P_0$, the joint $SPA$ system is in a pure state $|\phi\rangle_S \otimes |\Phi_G\rangle_{PA}$, where $|\Phi_G\rangle$ is a TMSV state with gain $G$. Equation (3) indicates that the EA transduction efficiency $\eta_{\mathrm{EA}} = \eta/[\eta + (1-\eta)/G]$ approaches unity at the $G \gg 1$ limit, thus the output state of $P^{\mathrm{out}}$ after the two-mode anti-squeezer is in the state $|\phi\rangle_{P^{\mathrm{out}}}$ identical to the input signal and decoupled from $S^{\mathrm{out}}A^{\mathrm{out}}$. Because the nonlinear coupler and anti-squeezer are unitary when $\kappa_E = 0$, the final joint state of $S^{\mathrm{out}}P^{\mathrm{out}}A^{\mathrm{out}}$ is pure, and thus $S^{\mathrm{out}}A^{\mathrm{out}}$ is pure. Therefore, at the $G \gg 1$ limit, the joint state equals $|\phi\rangle_{P^{\mathrm{out}}} \otimes |\Phi_{\tilde{G}}\rangle_{S^{\mathrm{out}}A^{\mathrm{out}}}$, where $S^{\mathrm{out}}A^{\mathrm{out}}$ is in an entangled TMSV. For finite $G$ value, derivations based on covariance matrix for vacuum input of $S$ confirm that the output state of $S^{\mathrm{out}}A^{\mathrm{out}}$ is a pure TMSV with the photon number:

$$N_S = \eta(G-1). \quad (7)$$

## 7. DUAL-BAND COOPERATIVE ENTANGLEMENT-ASSISTED PROTOCOL

A natural follow-up is to utilize the transduction output $P^{\mathrm{out}}$ also for distributing entanglement, besides the entanglement being generated between $S^{\mathrm{out}}A^{\mathrm{out}}$. To describe the entanglement source, we introduce a reference system $B$, entangled with $S$ of two-mode squeezing gain $G_S$, as indicated in Fig. 2. When $G_S \ll \eta G$, the EA transduction perfectly transduces the entanglement source according to Eq. (3) given $G \gg 1$, but at a negligible rate compared to Eq. (7). However, the transduction of strong entanglement with $G_S$ comparable to $\eta G$ is no longer perfect at $G \gg 1$: the reflected signal system $S'$ is now significantly correlated with $B$, which precludes the desired maximal entanglement across $P^{\mathrm{out}}B^{\mathrm{out}}$ and across $S^{\mathrm{out}}A^{\mathrm{out}}$ due to monogamy of entanglement.

To address this issue, we introduce the dual-band cooperative EA protocol. As shown in Fig. 2, now we anti-squeeze both bands to decouple the two entangled pairs: $S^{\mathrm{out}}A^{\mathrm{out}}$ and $P^{\mathrm{out}}B^{\mathrm{out}}$. The anti-squeezing gains $G'_S$ and $G'$ are optimized such that the local correlations within $PA$ and $SB$ are eliminated. Below we present our major findings in a symmetric and lossless case, while full derivations allowing coupling loss and ancilla loss are given in Appendix B.

In the symmetric case, the reflectivity $\kappa$, intrinsic loss $\kappa_E$, and squeezing gain $G$ are identical for both $S$ and $P$. Under lossless condition $\kappa_E = 0$, we show that pure TMSV pairs are generated for both the $S^{\mathrm{out}}A^{\mathrm{out}}$ and $P^{\mathrm{out}}B^{\mathrm{out}}$ individually, with the mean photon number of each mode:

$$N_S = \frac{1}{2}\left(\sqrt{4\eta(G-1)G+1} - 1\right). \quad (8)$$

For strong squeezing $G \gg 1$, the above $N_S \simeq \sqrt{\eta}G$. Compared with the single-band squeezed case of Eq. (7), entangled mean photon number $N_S$ is improved by a factor of $1/\sqrt{\eta}$.

The physical origin of such improved scaling with $\eta$ is the coherent interference between the two transduction paths, which has demonstrated non-trivial advantages beyond the direct summation [48]. Here, the cooperation of the two inputs in our dual-band protocols leads to two amplifications for each channel use, compared with only one amplification in the single-band protocol.

## 8. ANCILLA-FREE PROTOCOL

In fact, with dual-band cooperation and environment control, a beamsplitter can directly generate entanglement between the two bands without any ancilla assistance. For example, a balanced beamsplitter with two single-mode squeezed-vacuum state inputs, with squeezed quadratures chosen to be orthogonal, generates an ideal TMSV entanglement between the two output ports, and the entanglement strength increases with the input squeezing strength.

To distinguish such environment control advantage from the EA advantage, we introduce the ancilla-free protocol, as shown in Fig. 3. Compared with the EA protocol Fig. 2, the ancillae are removed, and the two-mode squeezers and anti-squeezers are replaced with their single-mode version, i.e., the input modes are degenerate: for squeezers $\hat{\mathcal{E}}_P = \sqrt{G}\hat{\mathcal{E}}_{P_0} + \sqrt{G-1}\hat{\mathcal{E}}^{\dagger}_{P_0}$, $\hat{\mathcal{E}}_S = \sqrt{G_S}\hat{\mathcal{E}}_{S_0} - \sqrt{G_S-1}\hat{\mathcal{E}}^{\dagger}_{S_0}$, for anti-squeezers $\hat{\mathcal{E}}_{P^{\rm out}} = \sqrt{G'}\hat{\mathcal{E}}_{P'} - \sqrt{G'-1}\hat{\mathcal{E}}^{\dagger}_{P'}$, $\hat{\mathcal{E}}_{S^{\rm out}} = \sqrt{G'_S}\hat{\mathcal{E}}_{S'} + \sqrt{G'_S-1}\hat{\mathcal{E}}^{\dagger}_{S'}$. Note that the phases for signal squeezer and anti-squeezer are different by $\pi$ from the probe to generate entanglement. For the evaluation of EPR squeezing $\Delta^{-}_{\rm EPR}$, we numerically optimize $G', G'_S$ to minimize $\Delta^{-}_{\rm EPR}$; for the evaluation of entanglement entropy and logarithmic negativity, these quantities are unchanged for any $G', G'_S$, since anti-squeezing is a local unitary transformation.

In the symmetric case with lossless coupling, we find a pure TMSV pair can be generated via proper anti-squeezing, with the mean photon number of each mode:

$$N_S = \frac{1}{2}\left(\sqrt{1+16(1-\eta)\eta(G-1)G} - 1\right). \tag{9}$$

For strong squeezing $G \gg 1$, $N_S \simeq 2\sqrt{(1-\eta)\eta}G$. The generated crossband entanglement per TMSV pair is even stronger than the dual-band EA protocol; however the dual-band EA protocol generates two pairs per channel use, and the ebit rate scales as $E \sim \log(N_S)$, hence the dual-band EA protocol is superior by a factor of 2 in terms of the total ebit rate scaling with $G$. Here we define one channel use as one use of the cross-frequency-band nonlinear coupling, allowing dual-band cooperation and entanglement assistance in general.

## 9. IDEAL LOSSLESS SYSTEM PERFORMANCE

In Fig. 4, we compare the performance of the single-band (blue), dual-band (red) entanglement-assisted protocols and the ancilla-free (brown) protocol, quantified by EPR quadrature squeezing $\Delta^{-}_{\rm EPR}$ in subplot (a) and entanglement entropy $E$ in subplot (b).

Both EA protocols demonstrate advantage $\propto G$ for $\Delta^{-}_{\rm EPR}$ and $2^E$ over the baseline of unassisted protocols. Furthermore, the dual-band cooperative protocol further improves the single-band EA protocol by $\sim\sqrt{\eta} = 10$ dB advantage in EPR squeezing [Subplot (a), 8.38 dB is achieved at 20 dB], as predicted by Eq. (8). We observe that merely 3 dB squeezing is sufficient for both our EA proposals to beat the baseline. In addition, we point out that while the single-band protocol generates a single pair of crossband entanglement, the dual-band protocol generates two pairs simultaneously.

Now consider the ancilla-free protocol. In subplot (a), the ancilla-free protocol yields an even stronger EPR squeezing than both the EA protocols, 3 dB stronger than the dual-band EA, as predicted by Eq. (9). Nevertheless, the dual-band EA protocol produces two crossband entanglement pairs per channel use. In subplot (b), we evaluate the sum of entanglement entropies $E$ over the two pairs for the dual-band EA protocol, which yields a two-fold advantage over the ancilla-free protocol when input squeezing gain $G$ is high.

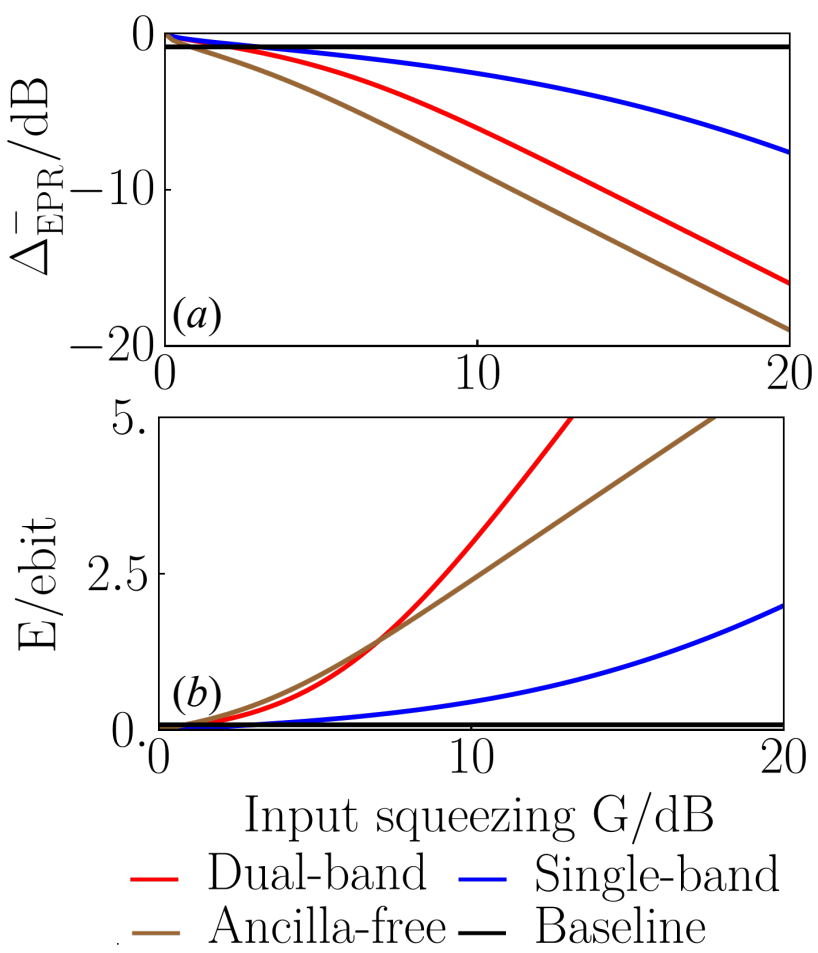

**Fig. 4.** The performance of crossband entanglement generation versus input two-mode squeezing gain $G$, under intrinsic transduction efficiency $\eta = 1$ % and coupling lossless limit $\kappa_E \to 0$. (a) EPR squeezing $\Delta^{-}_{\rm EPR}$/dB; (b) entanglement entropy $E$/ebit. Red: Dual-band cooperative EA entanglement generation with symmetric squeezing gain $G_S = G$ at both bands, $\Delta^{-}_{\rm EPR}$ evaluated across $S^{\rm out}A^{\rm out}$ while $E$ evaluated as the two-pair sum over $S^{\rm out}A^{\rm out}$ and $P^{\rm out}B^{\rm out}$; Blue: single-band EA transduction and entanglement generation with only $PA$ band squeezed by $G$, while $SB$ band $G_S = 1$; Brown: ancilla-free protocol; Black: baseline benchmark Eq. (6). We assume lossless ancilla.

## 10. PRACTICAL SYSTEM PERFORMANCE

So far we have assumed lossless coupling ($\kappa_E = 0$) in our analysis. Now we discuss the effect of $\kappa_E > 0$ in Eq. (1).

### A. Symmetric Case

To obtain an intuitive physical picture, first we assume the cavity and squeezing symmetric for both bands, i.e., the reflectivity $\kappa$, intrinsic loss $\kappa_E$, and input squeezing gain $G$ are identical for both $S$ and $P$.

For the single-band protocol, we can obtain the output covariance matrix of $S^{\rm out}A^{\rm out}$ when the input signal port $S$ is trivially in a vacuum state:

$$V^{SA}_{\rm out} = \begin{pmatrix} 2\gamma N_S + 1 & 0 & -2c_p & 0 \\ 0 & 2\gamma N_S + 1 & 0 & 2c_p \\ -2c_p & 0 & 2N_S + 1 & 0 \\ 0 & 2c_p & 0 & 2N_S + 1 \end{pmatrix}, \tag{10}$$

where $c_p \equiv \sqrt{\gamma N_S(1+N_S)}$, $N_S = (G-1)(1-\kappa)$ and $\gamma = \eta/(1-\kappa)$. Here the output can be viewed as a lossy TMSV—produced from a pure TMSV with mean photon number $N_S$ and $S$ going through a bosonic loss channel of transmissivity $\gamma$.

For the dual-band protocol, we find that two lossy TMSV pairs can be achieved for $S^{\rm out}A^{\rm out}$ and $P^{\rm out}B^{\rm out}$ individually, with covariance matrix in the same form of Eq. (10) but with different $N_S, \gamma$. The full analysis is presented in Appendix B.1, while we find at the strong squeezing limit $G \to \infty, N_S = G\left(\chi + \frac{\kappa_E}{2}\right)$ $\gamma = 1 - \frac{\kappa_E}{\chi + \kappa_E/2}$ where $\chi = \sqrt{\kappa_E^2/4 + \eta}$.

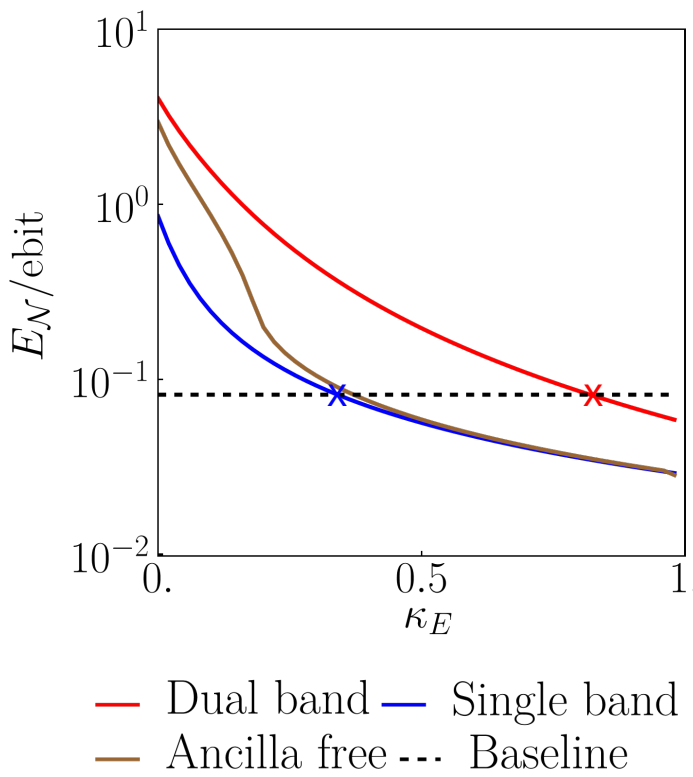

**Fig. 5.** Logarithmic negativity $E_{\mathcal{N}}$ across the *SA* output ports versus intrinsic loss $\kappa_E$, under optimized input two-mode squeezing gain $G$ bounded by $G \leq 10$. Intrinsic transduction efficiency is $\eta = 1$ %. Red solid: Dual-band cooperative EA entanglement generation with symmetric squeezing gain $G_S = G$ at both bands; Blue solid: single-band EA transduction and entanglement generation, with only *PA* band squeezed by $G$, while *SB* band $G_S = 1$; Brown solid: ancilla-free protocol with symmetric single-mode squeezing $G_S = G$; Black dashed: lossless baseline benchmark Eq. (6); Cross: loss threshold to achieve any advantage. We assume lossless ancilla.

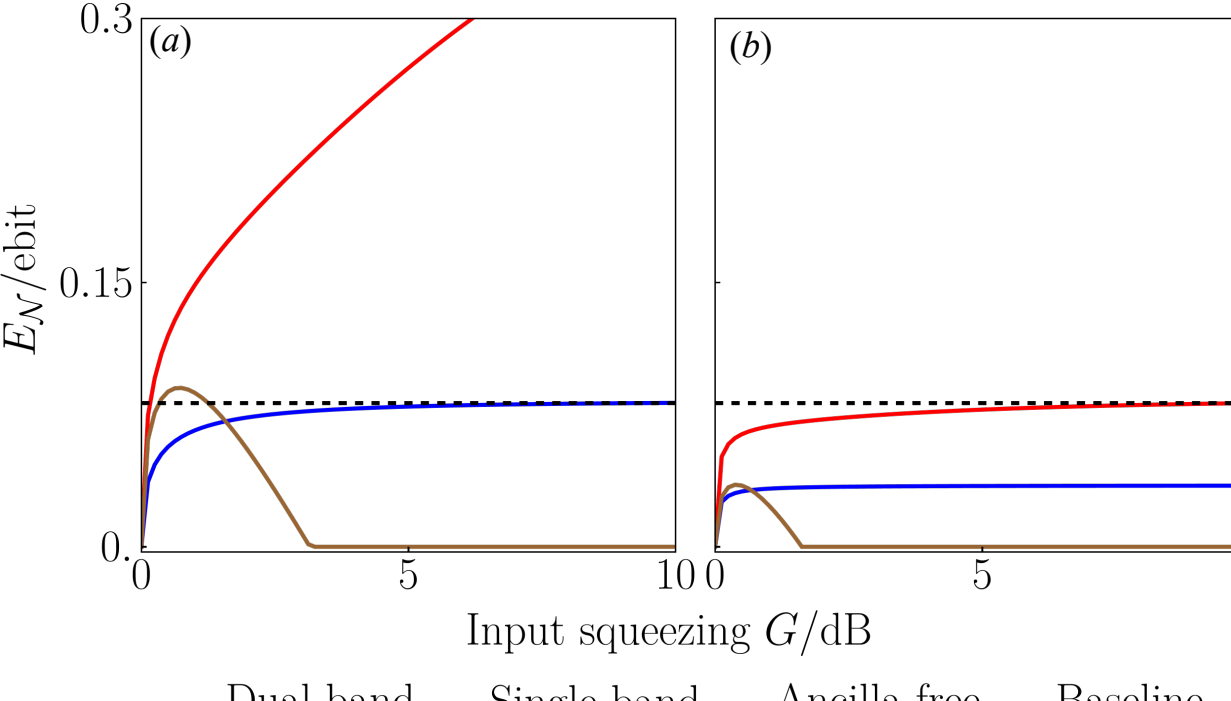

**Fig. 6.** Logarithmic negativity $E_{\mathcal{N}}$ across the *SA* output ports versus input two-mode squeezing gain $G$, in the presence of high intrinsic loss (a) $\kappa_E = 34$ %; (b) $\kappa_E = 82$ %. Intrinsic transduction efficiency is $\eta = 1$ %. Red: Dual-band cooperative EA entanglement generation with symmetric squeezing gain $G_S = G$ at both bands; Blue: single-band EA transduction and entanglement generation, with only *PA* band squeezed by $G$, while *SB* band $G_S = 1$; Brown: ancilla-free protocol with symmetric single-mode squeezing $G_S = G$; Black dashed: lossless baseline benchmark Eq. (6). We assume lossless ancilla.

As now the output is in a mixed state, the entanglement entropy is no longer a proper measure and we evaluate the logarithmic negativity $E_{\mathcal{N}}$ (see Appendix A) versus coupling loss $\kappa_E$ in Fig. 5, to explore the loss robustness, given unassisted efficiency $\eta = 1$%. Overall, we observe that the advantages of all the three protocols we proposed are robust to loss, especially the dual-band EA protocol (red) which combines the EA advantage and the environment control advantage. Given a realistic loss $\kappa_E = 5$%, the dual-band EA protocol yields 2.38 ebits, a 29-fold advantage over the baseline 0.082 ebits. We define $\kappa_E$ value at the cross point of our protocols with the baseline as the loss threshold (marked by crosses), which is the maximum tolerable loss that the protocol can achieve any advantage over the baseline due to the monotonicity with $\kappa_E$. From numerical evaluation at $\eta = 1$ %, we obtain the loss thresholds 34 % for the single-band EA protocol and 82 % for the dual-band EA protocol.

In Fig. 6, we evaluate the logarithmic negativity versus the input squeezing $G$ under the loss thresholds $\kappa_E = 34$ %, 82 % for the single-band and dual-band EA protocols, respectively. In subplot (a) with smaller loss $\kappa_E = 34$ %, we observe that the dual-band EA protocol (red) yields entanglement $E_{\mathcal{N}}$ scaling up with $G$, while $E_{\mathcal{N}}$ of single-band EA protocol (blue) saturates at merely 3 dB squeezing but it still allows an advantage over the lossless baseline benchmark (black dashed). In subplot (b) with a higher loss $\kappa_E = 82$ %, the dual-band protocol still yields advantage over the baseline. Its advantage over the single-band protocol converges to two folds, owing to the two output entanglement pairs. In contrast, in both cases $E_{\mathcal{N}}$ of ancilla-free protocol (brown) diminishes at strong squeezing limit, the maximal $E_{\mathcal{N}}$ values are significantly worse than the EA protocols, especially for an intermediate loss in subplot (a). For the strong loss in subplot (b), the optimal performance of ancilla-free protocol is almost identical to the single-band EA protocol. Considering the regime of strong input squeezing, where the input squeezing engineering does not depend on the unassisted efficiency $\eta$ of the coupling system, the entanglement assistance is required to maintain the advantage over the direct-transduction limit in the presence of strong loss. This is useful for broadband entanglement generation, where engineering of input squeezing spectrum $G(\omega)$ according to the $\eta(\omega)$ spectrum, e.g. a Lorentzian shape for a cavity, can be highly challenging.

## B. General Case

Now we consider a realistic experiment parameter setup of cavity microwave-optical quantum transduction from the high cooperativity setup in Ref. [49]. Since the dual-band EA protocol has been demonstrated with the best loss-robust crossband entanglement generation performance, here we focus on the dual-band EA protocol only. For simplicity we consider the signal mode and probe mode at the on-resonance frequency only. To highlight the quantum advantage, we consider the overcoupled cavity with high external coupling efficiency $\zeta_o = 95$ % (optical), $\zeta_e = 99$ % (microwave), which invoke optical loss $\propto 1 - \zeta_o$ and microwave loss $\propto 1 - \zeta_e$. We consider microwave signal and optical probe as an example; the transduction is bidirectional and the labels of signal $S$ and probe $P$ are interchangeable in general. The numerical evaluation results of logarithmic negativity $E_{\mathcal{N}}$ of the crossband entanglement are shown in Fig. 7. To reveal the structural features from the asymmetry of microwave and optical losses, here we plot the output crossband entanglement at $P^{\text{out}}B^{\text{out}}$ and $S^{\text{out}}A^{\text{out}}$ separately. Note that for $P^{\text{out}}B^{\text{out}}$, the microwave squeezing is the entanglement source, the optical squeezing is the entanglement assistance; for $S^{\text{out}}A^{\text{out}}$, vice versa.

In subplots (a1,a2) of Fig. 7, we show the evaluation given a high cooperativity $C = 0.38$, the setup in Ref. [49] with noise photon demonstrated below 1. First let us consider the $P^{\text{out}}B^{\text{out}}$ output in subplot (a1). We observe that the output crossband entanglement increases (from bright to dark) as entanglement source strength $G_S$ increases. Meanwhile, the output crossband entanglement never increases with the entanglement assistance strength $G$; this is because the optical entanglement assistance invokes more noises due to the optical loss larger than the microwave loss. We note that this is a novel phenomenon never found in the original EA transduction [37]. Compared with the baseline benchmark 2.8 ebits (red), we observe quantum advantage in the presence of cavity loss, even though the baseline benchmark assumes no loss. A similar phenomenon occurs for the $S^{\text{out}}A^{\text{out}}$ output in subplot (a2), with entanglement source to be transduced and entanglement assistance swapped as $G, G_S$, respectively, while the main difference is that, here the output crossband entanglement increases with the microwave entanglement assistance strength $G_S$, because the microwave loss is weaker. Therefore, here our dual-band EA protocol yields a larger quantum advantage over the baseline benchmark (red), always increasing with $G_S$. In contrast, the optical entanglement source $G$ now has an optimum, beyond which the output entanglement begins to decay, because the optical loss is stronger. Such monotonicity with $G_S$ and concavity with $G$ form the semi-ellipses at the larger $G_S$ side of the contour.

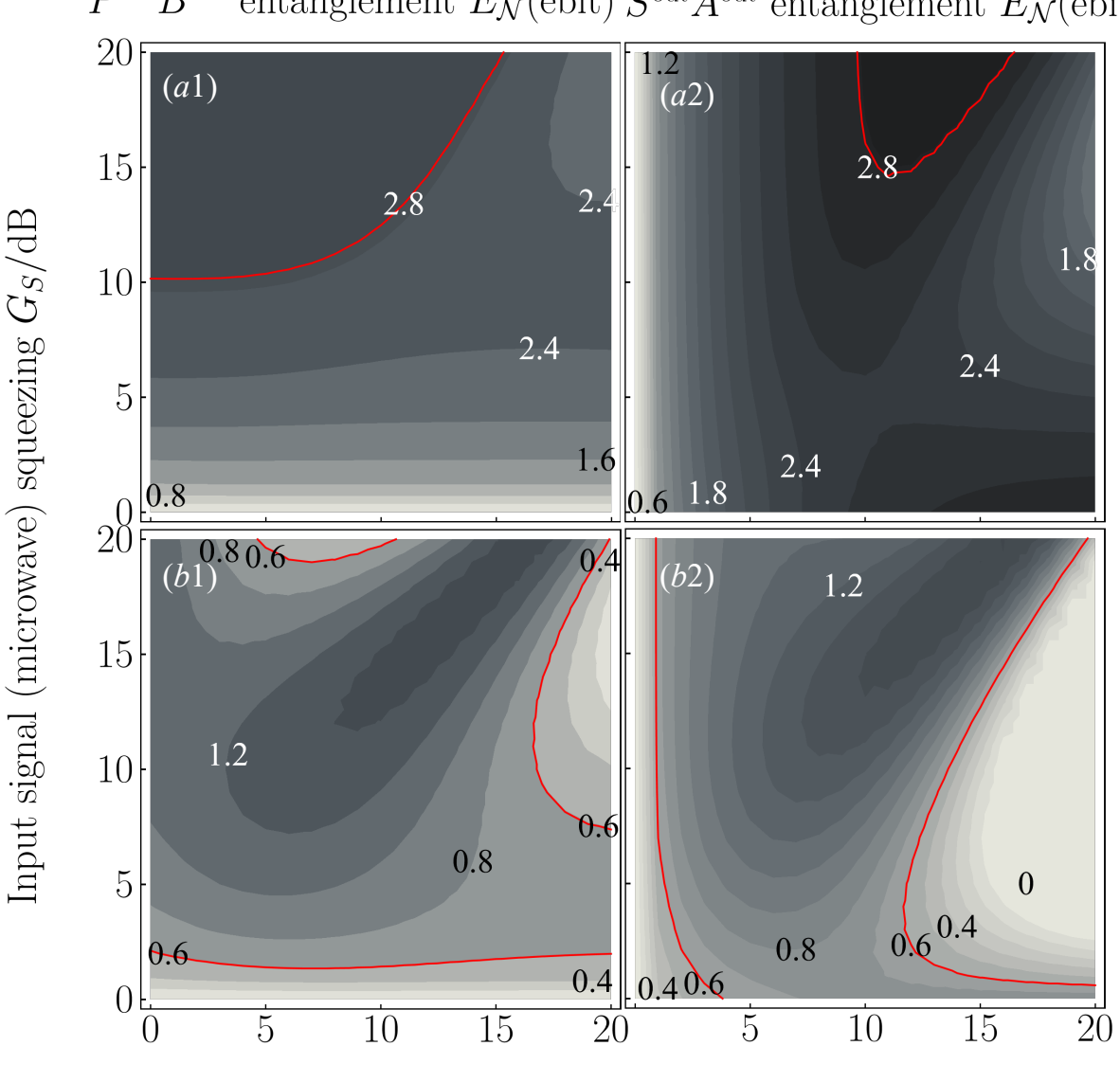

**Fig. 7.** Logarithmic negativity $E_{\mathcal{N}}$ (ebit) of the output entanglement from the dual-band EA protocol using an asymmetric cavity at the on-resonance frequency, across the crossband output pair $P^{\text{out}}B^{\text{out}}$ (subplots a1, b1) and $S^{\text{out}}A^{\text{out}}$ (subplots a2, b2), respectively. (a1) and (a2) Higher cooperativity $C = 0.38$; (b1) and (b2) Lower cooperativity $C = 0.038$. The red contour marks up the baseline benchmark. Cavity parameter chosen from the high cooperativity setup in Ref. [49], assuming microwave signal and optical probe, except for higher external coupling efficiency $\zeta_o = 95$ % (optical), $\zeta_e = 99$ % (microwave). We assume lossless ancilla.

In subplots (b1,b2) of Fig. 7, we decrease the cooperativity to $C = 0.038$. Now we observe the concavity holds for both the signal and probe, thus a complete ellipse appears. Here with smaller cooperativity, the baseline benchmark (red) is limited to merely 0.6 ebits, thus the quantum advantage of our protocol becomes more significant and much easier to achieve even with a weak input squeezing less than 3 dB. Here the quantum advantage increases with both $G, G_S$; this can be explained by the EA transducer theory Eq. (3) that the EA amplification of transduction efficiency in the plotted region is free from the saturation as $\eta_{\text{EA}} \to 100$ % given small $C$.

Overall, our dual-band EA protocol yields significant advantage over the baseline for practical cavities with high external coupling efficiency. When the cooperativity is smaller, the advantage can be achieved more easily by smaller input squeezing. Different from the symmetric case, a realistic asymmetric experiment system may require asymmetric input squeezing to optimize the entanglement generation. And the optimization remains an open problem since the optimum of $(G, G_S)$ to maximize $P^{\text{out}}B^{\text{out}}$ entanglement is in general not identical to that maximizing $S^{\text{out}}A^{\text{out}}$ entanglement, one needs to make a joint optimization to maximize the utility function, e.g., a weighted average of the two entanglement strengths, according to specific experiment setups.

## 11. DISCUSSION

Here we compare our findings with previous works. Lami *et al.* [50] and Mele *et al.* [51] proposed using Fock state environment assistance to achieve non-zero quantum capacity, but an explicit encoder–decoder construction of a capacity-achieving protocol is missing. Wang and Jiang [36] proposed to use environment assistance in the Gottesman–Kitaev–Preskill (GKP) state, but the transduced input requires specific GKP encoding, and the scaling of crossband entanglement generation rate with respect to the source brightness remains unknown. With intraband two-mode squeezers as the encoder and decoder, our proposal, to our knowledge, is the first explicit protocol achieving crossband entanglement generation rate beyond the vacuum-environment limit.

## APPENDIX A: ENTANGLEMENT MEASURE

The logarithmic negativity is defined as:

$$E_{\mathcal{N}}(\hat{\rho}) = \log \|\hat{\rho}\|, \tag{A1}$$

where $\|\hat{\rho}\| = \mathrm{Tr}\{|\hat{\rho}|\}$ is the trace norm. It was first proposed in Ref. [52], and later rigorously proven as a monotonic entanglement measure in Ref. [43]. It quantifies the maximum violation of the separability criterion for two-mode Gaussian states [44,45], and is an upper bound of the distillable entanglement of the quantum state [53].

For a two-mode Gaussian state, the logarithmic negativity is defined as [53]:

$$E_{\mathcal{N}}(\hat{\rho}) = \max\{0, -\log \nu\}, \tag{A2}$$

where $\nu$ is the smallest symplectic value of the partially transposed state.

As an example, a pair of modes $\hat{a}_S, \hat{a}_B$ in a TMSV state is generated from vacuum modes $\hat{a}_{S_0}, \hat{a}_{B_0}$ by the two-mode squeezing operation $\mathcal{S}(G_S) : \hat{a}_{S_0}, \hat{a}_{B_0} \to \hat{a}_S, \hat{a}_B$ with:

$$\begin{aligned} \hat{a}_S &= \sqrt{G_S}\hat{a}_{S_0} + \sqrt{G_S - 1}\hat{a}^\dagger_{B_0}, \\ \hat{a}_B &= \sqrt{G_S}\hat{a}_{B_0} + \sqrt{G_S - 1}\hat{a}^\dagger_{S_0}. \end{aligned} \tag{A3}$$

It is a Gaussian state with zero mean and covariance matrix:

$$\Lambda_{\text{TMSV}} = \begin{pmatrix} (2N_S + 1)\mathbf{I} & 2C_0\mathbf{Z} \\ 2C_0\mathbf{Z} & (2N_S + 1)\mathbf{I} \end{pmatrix}, \tag{A4}$$

where $\mathbf{I}$, $\mathbf{Z}$ are two-by-two Pauli matrices, $C_0 = \sqrt{N_S\,(N_S+1)}$ is the amplitude of the phase-sensitive cross correlation and $N_S = G_S - 1$ is the mean photon number. The variances of its EPR-type quadratures $\hat{q}_- \equiv \mathrm{Re}(\hat{a}_S - \hat{a}_B)$, $\hat{p}_+ \equiv \mathrm{Im}(\hat{a}_S + \hat{a}_B)$ are squeezed below the vacuum fluctuation. For such EPR-type entanglement, a natural entanglement measure is the average variance of the EPR quadratures $\hat{q}_-$ and $\hat{p}_+$, defined as $\Delta^-_{\mathrm{EPR}} \equiv (\mathrm{var}\;\hat{q}_- + \mathrm{var}\;\hat{p}_+)/2 = \frac{1}{2} + G_S - 1 - \sqrt{(G_S-1)\,G_S}$. At the strong squeezing limit $G_S \to \infty$, $\Delta^-_{\mathrm{EPR}} \simeq 1/8G_S$.

## APPENDIX B: DERIVATION OF THE FULL DUAL-BAND INPUT–OUTPUT RELATION

Here we derive the full input–output relation including all outputs $S^{\mathrm{out}}, B^{\mathrm{out}}, P^{\mathrm{out}}, A^{\mathrm{out}}$ of both bands.

We initialize the probe and ancilla modes $\hat{\mathcal{E}}_{P_0}, \hat{\mathcal{E}}_{A_0}$ in vacuum, and similarly for the signal mode and its ancilla $\hat{\mathcal{E}}_{S_0}, \hat{\mathcal{E}}_{B_0}$. Then we apply two-mode squeezers before nonlinear coupling, which gives:

$$
\begin{aligned}
\hat{\mathcal{E}}_P &= \sqrt{G}\hat{\mathcal{E}}_{P_0} + \sqrt{G-1}\hat{\mathcal{E}}^\dagger_{A_0}\,,\\
\hat{\mathcal{E}}_A &= \sqrt{G-1}\hat{\mathcal{E}}^\dagger_{P_0} + \sqrt{G}\hat{\mathcal{E}}_{A_0}\,,
\end{aligned}
\tag{B1}
$$

and

$$
\begin{aligned}
\hat{\mathcal{E}}_S &= \sqrt{G_S}\hat{\mathcal{E}}_{S_0} + \sqrt{G_S-1}\hat{\mathcal{E}}^\dagger_{B_0}\\
\hat{\mathcal{E}}_B &= \sqrt{G_S-1}\hat{\mathcal{E}}^\dagger_{S_0} + \sqrt{G_S}\hat{\mathcal{E}}_{B_0}\,.
\end{aligned}
\tag{B2}
$$

The nonlinear coupling of the transduction device forms a beamsplitter between the squeezed probe $\hat{\mathcal{E}}_P$ and the signal $\hat{\mathcal{E}}_S$:

$$
\begin{aligned}
\hat{\mathcal{E}}_{P'} &= \sqrt{\kappa_P}\hat{\mathcal{E}}_P + \sqrt{\eta}\hat{\mathcal{E}}_S + \sqrt{\kappa_E}\hat{\mathcal{E}}_E,\\
\hat{\mathcal{E}}_{S'} &= \sqrt{\kappa_S}\hat{\mathcal{E}}_S - \sqrt{\eta}\hat{\mathcal{E}}_P + \frac{\sqrt{\kappa_P}-\sqrt{\kappa_S}}{|\sqrt{\kappa_P}-\sqrt{\kappa_S}|}\sqrt{\kappa^S_E}\hat{\mathcal{E}}_E + \sqrt{\kappa^S_F}\hat{\mathcal{E}}_F,
\end{aligned}
\tag{B3}
$$

where $\kappa^S_E = \frac{\eta}{\kappa_E}(\sqrt{\kappa_P} - \sqrt{\kappa_S})^2$, $\kappa^S_F = 1 - \eta - \kappa_S - \kappa^S_E$. The intrinsic losses are $\kappa_E = 1 - \eta - \kappa_P$, $\kappa_F = \kappa^S_E + \kappa^S_F = 1 - \eta - \kappa_S$ for the probe mode and signal mode, respectively, and the environment modes $\hat{\mathcal{E}}_E, \hat{\mathcal{E}}_F$ are in vacuum, we omit the phase shifts for simplicity. Here the formula of $\hat{\mathcal{E}}_{S'}$ is more complicated because we define the modes to make $\hat{\mathcal{E}}_{P'}$ simple, and then obtain $\hat{\mathcal{E}}_{S'}$ such that the transformation is symplectic. Alternatively, one can choose the opposite way to obtain a simple formula for $\hat{\mathcal{E}}_{S'}$ at the cost of complication of $\hat{\mathcal{E}}_{P'}$.

Meanwhile ancillae $\hat{\mathcal{E}}_A$, $\hat{\mathcal{E}}_B$ are stored in quantum memory with transmissivity $\kappa_A, \kappa_B$. In the main text, we focus on the symmetric case $\kappa = \kappa_S$, $\kappa_E = \kappa_F$, $\kappa_A = \kappa_B$, thus $\kappa, \kappa_E, \kappa_A$ alone are sufficient to define the coupling reflectivity, intrinsic loss, ancilla loss. In the main text we also assume ancilla lossless $\kappa_A = 1$, since the storage quantum memory can possess a much higher lifetime while the probe goes through a lossy low-lifetime nonlinear coupling with the signal. Below, we derive for the general asymmetric case, with numerical evaluations presented in Appendix B.2.

After the nonlinear coupling at the transduction device, we apply the anti-squeezers for both bands:

$$
\begin{aligned}
\hat{\mathcal{E}}_{P^{\mathrm{out}}} &= \sqrt{G'}\hat{\mathcal{E}}_{P'} - \sqrt{G'-1}\hat{\mathcal{E}}^\dagger_A,\\
\hat{\mathcal{E}}_{A^{\mathrm{out}}} &= -\sqrt{G'-1}\hat{\mathcal{E}}^\dagger_{P'} + \sqrt{G'}\hat{\mathcal{E}}_A\,,
\end{aligned}
\tag{B4}
$$

and

$$
\begin{aligned}
\hat{\mathcal{E}}_{S^{\mathrm{out}}} &= \sqrt{G'_S}\hat{\mathcal{E}}_{S'} - \sqrt{G'_S-1}\hat{\mathcal{E}}^\dagger_B,\\
\hat{\mathcal{E}}_{B^{\mathrm{out}}} &= -\sqrt{G'_S-1}\hat{\mathcal{E}}^\dagger_{S'} + \sqrt{G'_S}\hat{\mathcal{E}}_B\,.
\end{aligned}
\tag{B5}
$$

To quantify the entanglement of a Gaussian state, first we define the field quadrature operator vector $\hat{\boldsymbol{x}} \equiv [\hat{X}_B, \hat{Y}_B, \hat{X}_S, \hat{Y}_S, \hat{X}_P, \hat{Y}_P, \hat{X}_A, \hat{Y}_A]^T$, where $\hat{X}_\cdot = \hat{\mathcal{E}}_\cdot + \hat{\mathcal{E}}^\dagger_\cdot$, $\hat{Y}_\cdot = (\hat{\mathcal{E}}_\cdot - \hat{\mathcal{E}}^\dagger_\cdot)/i$, and its covariance matrix $V_{jk} \equiv \langle\{\Delta\hat{x}_j, \Delta\hat{x}_k\}\rangle/2$, where $\{\hat{a}, \hat{b}\} = \hat{a}\hat{b} + \hat{b}\hat{a}$ is the anti-commutator, $\Delta\hat{x} \equiv \hat{x} - \langle\hat{x}\rangle$. We consider a single mode of the fields for simplicity, which satisfies the canonical commutation relation $[\hat{X}_\cdot, \hat{Y}_\cdot] = 2i$. Given a finite squeezing gain $G$ and an anti-squeezing gain $G'$, the full formula of the output covariance matrix is complicated, we present it in Appendix B.2. We choose the anti-squeezing gains $G', G'_S$ to decouple $P$ from $A$ to purify the output entanglement, the formula is in Appendix B.2 alongside.

Below we first present the formula of the covariance matrix of the output quantum state for the cases of symmetric squeezing with lossless ancilla in Appendix B.1, then the full formulas of general cases in Appendix B.2. An analysis of single-band entanglement assistance $G_S = 1$ is in Appendix C.

### B.1. Symmetric Squeezing with Lossless Ancilla

For symmetric squeezing $G = G_S$ with symmetric coupling reflectivity $\kappa_S = \kappa$ (thus $\kappa_E = \kappa_F = 1 - \eta - \kappa$), and lossless ancillas $\kappa_A = \kappa_B = 1$, we have the output covariance matrix:

$$
V_{\mathrm{lossless}} = \begin{pmatrix}
2N_S+1 & 0 & 0 & 0 & 2c_p & 0 & 0 & 0\\
0 & 2N_S+1 & 0 & 0 & 0 & -2c_p & 0 & 0\\
0 & 0 & 2\gamma N_S+1 & 0 & 0 & 0 & -2c_p & 0\\
0 & 0 & 0 & 2\gamma N_S+1 & 0 & 0 & 0 & 2c_p\\
2c_p & 0 & 0 & 0 & 2\gamma N_S+1 & 0 & 0 & 0\\
0 & -2c_p & 0 & 0 & 0 & 2\gamma N_S+1 & 0 & 0\\
0 & 0 & -2c_p & 0 & 0 & 0 & 2N_S+1 & 0\\
0 & 0 & 0 & 2c_p & 0 & 0 & 0 & 2N_S+1
\end{pmatrix},
\tag{B6}
$$

where $c_p \equiv \sqrt{\gamma N_S(1+N_S)}$, and:

$$
\begin{aligned}
N_S &= \frac{1}{2}\left(\sqrt{\eta^2(G-1)^2 + 2\eta(G-1)((G-1)\kappa+G) + (G-(G-1)\kappa)^2} - \eta G + \eta + G - (G-1)\kappa - 2\right),\\
\gamma &= \frac{\sqrt{\eta^2(G-1)^2 + 2\eta(G-1)((G-1)\kappa+G) + (G-(G-1)\kappa)^2} + \eta(G-1) + G(\kappa-1) - \kappa}{\sqrt{\eta^2(G-1)^2 + 2\eta(G-1)((G-1)\kappa+G) + (G-(G-1)\kappa)^2} + \eta(-G) + \eta - G\kappa + G + \kappa - 2}.
\end{aligned}
\tag{B7}
$$

Here the output can be viewed as a tensor product of pure TMSV states with photon number $N_S$ after $S$ and $P$ go through bosonic loss channels of transmissivity $\gamma$.

At the limit of strong squeezing $G \to \infty$, with $\chi = \sqrt{\kappa_E^2/4 + \eta}$, we have:

$$N_S = G\left(\chi + \frac{\kappa_E}{2}\right), \; \gamma = 1 - \frac{\kappa_E}{\chi + \kappa_E/2}. \tag{B8}$$

### B.2. Full Formulas of the Output Covariance Matrix and Anti-Squeezing Gains

Here we present the full formulas of the output covariance matrix and anti-squeezing gains, for general cases. The full formula of the output covariance matrix is:

$$V = \begin{pmatrix} 2\nu_B+1 & 0 & 2c_{\rm SB} & 0 & 2c_{\rm PB} & 0 & 2c_{\rm AB} & 0 \\ 0 & 2\nu_B+1 & 0 & -2c_{\rm SB} & 0 & -2c_{\rm PB} & 0 & 2c_{\rm AB} \\ 2c_{\rm SB} & 0 & 2\nu_S+1 & 0 & 2c_{\rm PS} & 0 & 2c_{\rm SA} & 0 \\ 0 & -2c_{\rm SB} & 0 & 2\nu_S+1 & 0 & 2c_{\rm PS} & 0 & -2c_{\rm SA} \\ 2c_{\rm PB} & 0 & 2c_{\rm PS} & 0 & 2\nu_P+1 & 0 & 2c_{\rm PA} & 0 \\ 0 & -2c_{\rm PB} & 0 & 2c_{\rm PS} & 0 & 2\nu_P+1 & 0 & -2c_{\rm PA} \\ 2c_{\rm AB} & 0 & 2c_{\rm SA} & 0 & 2c_{\rm PA} & 0 & 2\nu_A+1 & 0 \\ 0 & 2c_{\rm AB} & 0 & -2c_{\rm SA} & 0 & -2c_{\rm PA} & 0 & 2\nu_A+1 \end{pmatrix}, \tag{B9}$$

where

$$\begin{aligned} \nu_B &= -G_S\kappa_S + (G_S-1)\,G'_S\kappa_S + \kappa_S + \eta(G-1)\left(G'_S-1\right) + (G_S-1)\,\kappa_B G'_S + G'_S - 2\sqrt{(G_S-1)\,G_S\kappa_B\kappa_S\left(G'_S-1\right)G'_S} - 1, \\ \nu_S &= \left(\sqrt{G_S\kappa_B\left(G'_S-1\right)} - \sqrt{(G_S-1)\,\kappa_S G'_S}\right)^2 + \kappa_B + \eta(G-1)G'_S - \kappa_B G'_S + G'_S - 1, \\ \nu_P &= \left(\sqrt{(G-1)\kappa G'} - \sqrt{G\kappa_A\left(G'-1\right)}\right)^2 + \kappa_A + \eta\,(G_S-1)\,G' - \kappa_A G' + G' - 1, \\ \nu_A &= \left(\sqrt{G\kappa\left(G'-1\right)} - \sqrt{(G-1)\kappa_A G'}\right)^2 - (\eta+\kappa-1)\left(G'-1\right) + \eta G_S\left(G'-1\right), \end{aligned} \tag{B10}$$

$$\begin{aligned} c_{\rm PS} &= -\sqrt{\eta\kappa G' G'_S}(G-1) - \sqrt{\eta\,(G_S-1)\,G_S\kappa_B G'\left(G'_S-1\right)} + \sqrt{\eta(G-1)G\kappa_A\left(G'-1\right)G'_S} + (G_S-1)\sqrt{\eta\kappa_S G' G'_S}, \\ c_{\rm PA} &= \sqrt{\left(G'-1\right)G'}(\eta+\kappa-1) + \left(\sqrt{G\kappa G'} - \sqrt{(G-1)\kappa_A\left(G'-1\right)}\right)\left(\sqrt{(G-1)\kappa_A G'} - \sqrt{G\kappa\left(G'-1\right)}\right) - \eta G_S\sqrt{\left(G'-1\right)G'}, \\ c_{\rm PB} &= -\sqrt{\eta\kappa_S G'\left(G'_S-1\right)}\,(G_S-1) + \left(\sqrt{(G-1)\kappa G'} - \sqrt{G\kappa_A\left(G'-1\right)}\right)\sqrt{\eta(G-1)\left(G'_S-1\right)} + \sqrt{\eta\,(G_S-1)\,G_S\kappa_B G' G'_S}, \\ c_{\rm SA} &= \sqrt{\eta\kappa\left(G'-1\right)G'_S}(G-1) + \sqrt{\eta\,(G_S-1)\left(G'-1\right)}\left(\sqrt{G_S\kappa_B\left(G'_S-1\right)} - \sqrt{(G_S-1)\,\kappa_S G'_S}\right) - \sqrt{\eta(G-1)G\kappa_A G' G'_S}, \\ c_{\rm SB} &= -\eta\sqrt{\left(G'_S-1\right)G'_S}(G-1) + \left(\sqrt{G_S\kappa_B G'_S} - \sqrt{(G_S-1)\,\kappa_S\left(G'_S-1\right)}\right)\left(\sqrt{(G_S-1)\,\kappa_S G'_S} - \sqrt{G_S\kappa_B\left(G'_S-1\right)}\right) \\ &\quad + (\kappa_B-1)\sqrt{\left(G'_S-1\right)G'_S}, \\ c_{\rm AB} &= -\sqrt{\eta\kappa\left(G'-1\right)\left(G'_S-1\right)}(G-1) + (G_S-1)\sqrt{\eta\kappa_S\left(G'-1\right)\left(G'_S-1\right)} + \sqrt{\eta(G-1)G\kappa_A G'\left(G'_S-1\right)} \\ &\quad - \sqrt{\eta\,(G_S-1)\,G_S\kappa_B\left(G'-1\right)G'_S}. \end{aligned} \tag{B11}$$

To decouple $P$ from $A$ to purify the output entanglement, we choose the anti-squeezing gains $G', G'_S$ as follows:

$$G'^\star = \frac{1}{2}\left(\frac{\eta\,(G_S-1) + G\kappa + G - \kappa}{\sqrt{\eta^2\,(G_S-1)^2 + 2\eta(G\kappa + G - \kappa)\,(G_S-1) + (G(1-\kappa)+\kappa)^2}} + 1\right), \tag{B12}$$

$$G_S'^\star = \frac{1}{2}\left(\frac{\eta(G-1) + G_S\kappa_S + G_S - \kappa_S}{\sqrt{\eta^2(G-1)^2 + 2\eta(G-1)\,(G_S\kappa_S + G_S - \kappa_S) + (G_S\,(1-\kappa_S)+\kappa_S)^2}} + 1\right), \tag{B13}$$

such that the intraband correlations within $PA$ and $SB$ respectively are eliminated when ancilla is lossless $\kappa_A, \kappa_B \to 1$. We note that nonzero unwanted crossband correlation between $P, S$ and between $A, B$ remains if the two bands are not symmetric, $\kappa \neq \kappa_S$ or $G \neq G_S$.

## APPENDIX C: SINGLE-BAND ENTANGLEMENT ASSISTANCE $G_S = 1$

Here we continue the analysis in Appendix B, assuming probe-band entanglement assistance with the signal band classical $G_S = 1$. With $G_S = 1$, at the $PA$ band, the anti-squeezer gives the

probe output:

$$
\begin{aligned}
\hat{\mathcal{E}}_{P^{\text{out}}} &= \sqrt{G'}\left(\sqrt{\kappa}\hat{\mathcal{E}}_P + \sqrt{\eta}\hat{\mathcal{E}}_S + \sqrt{\kappa_E}\hat{\mathcal{E}}_E\right) - \sqrt{G'-1}\hat{\mathcal{E}}_A^\dagger \\
&= \left(\sqrt{G\kappa G'} - \sqrt{(G'-1)(G-1)}\right)\hat{\mathcal{E}}_{P_0} \\
&\quad + \sqrt{\eta G'}\hat{\mathcal{E}}_S + \left(\sqrt{(G-1)\kappa G'} - \sqrt{(G'-1)G}\right)\hat{\mathcal{E}}_{A_0}^\dagger \\
&\quad + \sqrt{(1-\eta-\kappa)\,G'}\hat{\mathcal{E}}_E\,. \\
\hat{\mathcal{E}}_{A^{\text{out}}} &= -\sqrt{G'-1}\left(\sqrt{\kappa}\hat{\mathcal{E}}_P^\dagger + \sqrt{\eta}\hat{\mathcal{E}}_S^\dagger + \sqrt{\kappa_E}\hat{\mathcal{E}}_E^\dagger\right) + \sqrt{G'}\hat{\mathcal{E}}_A \\
&= \left(-\sqrt{G\kappa(G'-1)} + \sqrt{G'(G-1)}\right)\hat{\mathcal{E}}_{P_0}^\dagger \\
&\quad - \sqrt{\eta(G'-1)}\hat{\mathcal{E}}_S^\dagger \\
&\quad + \left(-\sqrt{(G-1)\kappa(G'-1)} + \sqrt{G'G}\right)\hat{\mathcal{E}}_{A_0} \\
&\quad - \sqrt{(1-\eta-\kappa)\,(G'-1)}\hat{\mathcal{E}}_E.
\end{aligned} \tag{C1}
$$

For single-band entanglement assistance we have $G_S = 1$, $\hat{\mathcal{E}}_S$ is in vacuum state, then $G_S'^{\star} = 1$, and the optimal $G'^{\star}$ reduces to the choice in EA transducer [37]:

$$
G'^{\star}|_{G_S=1} = \frac{1}{1-\kappa+\kappa/G}. \tag{C2}
$$

By such anti-squeezing, we can obtain the ancilla output:

$$
\begin{aligned}
\hat{\mathcal{E}}_{A^{\text{out}}}|_{G_S=1,G'=G'^{\star}} = \hat{\mathcal{E}}_{A_0}\sqrt{G(1-\kappa)+\kappa} \\
- \hat{\mathcal{E}}_E^\dagger\sqrt{\frac{(G-1)\kappa(1-\eta-\kappa)}{G(1-\kappa)+\kappa}} \\
- \hat{\mathcal{E}}_{P_0}^\dagger(\kappa-1)\sqrt{\frac{(G-1)G}{G(1-\kappa)+\kappa}} \\
- \hat{\mathcal{E}}_S^\dagger\sqrt{\frac{\eta(G-1)\kappa}{G(1-\kappa)+\kappa}}.
\end{aligned} \tag{C3}
$$

The output probe carries information about the input signal:

$$
\hat{\mathcal{E}}^{\star}_{P^{\text{out}}} = \frac{\sqrt{\kappa}\hat{\mathcal{E}}_{P_0} + \sqrt{\eta G}\hat{\mathcal{E}}_S + \sqrt{(1-\eta-\kappa)G}\hat{\mathcal{E}}_E}{\sqrt{G\,(1-\kappa)+\kappa}}\,. \tag{C4}
$$

Note that $\hat{\mathcal{E}}_{P_0}, \hat{\mathcal{E}}_E$ are in vacuum state, the output can be written as:

$$
\hat{\mathcal{E}}^{\star}_{P^{\text{out}}} = \sqrt{\eta_{\text{EA}}}\hat{\mathcal{E}}_S + \sqrt{1-\eta_{\text{EA}}}\hat{\mathcal{E}}_{\text{VAC}}\,, \tag{C5}
$$

where the noise background $\hat{\mathcal{E}}_{\text{VAC}}$ is in vacuum state, the EA transduction efficiency is:

$$
\eta_{\text{EA}} = \eta G' = \frac{\eta G}{G(1-\kappa)+\kappa}\,, \tag{C6}
$$

which recovers the EA transducer from $S$ to $P^{\text{out}}$ in Ref. [37].

It is noteworthy that entanglement is simultaneously generated between $S^{\text{out}}A^{\text{out}}$ if the loss is low. At the lossless limit of $\kappa_E = \kappa_F = 0$ ($\kappa_P = \kappa_S = 1-\eta$), after combining Eqs. (B3) and (C3) we observe that $S, A$ are maximally entangled at the strong squeezing limit $G \to \infty$:

$$
\begin{aligned}
\hat{\mathcal{E}}_{S^{\text{out}}} &\to \sqrt{\eta G}\hat{\mathcal{E}}_{P_0} + \sqrt{\eta G}\hat{\mathcal{E}}_{A_0}^\dagger, \\
\hat{\mathcal{E}}_{A^{\text{out}}} &\to \sqrt{\eta G}\hat{\mathcal{E}}_{A_0} + \sqrt{\eta G}\hat{\mathcal{E}}_{P_0}^\dagger.
\end{aligned} \tag{C7}
$$

## APPENDIX D: FULL FORMULAS OF THE ANCILLA-FREE PROTOCOL

Using an analysis similar to the input–output relation derivation in Appendix B, we derive the output covariance matrix for the ancilla-free protocol:

$$
V = \begin{pmatrix} 2d_S + 2\nu_S + 1 & 0 & 2(c_1+c_2) & 0 \\ 0 & -2d_S + 2\nu_S + 1 & 0 & 2c_2 - 2c_1 \\ 2(c_1+c_2) & 0 & 2d_P + 2\nu_P + 1 & 0 \\ 0 & 2c_2 - 2c_1 & 0 & -2d_P + 2\nu_P + 1 \end{pmatrix}, \tag{D1}
$$

where

$$
\begin{aligned}
\nu_S &= -G_S\kappa_S + 2G_SG_S'\kappa_S - 2G_S'\kappa_S - 2\sqrt{(G_S-1)\,G_S\,(G_S'-1)\,G_S'}\kappa_S + \kappa_S + G_S' + \eta\,(-2G_S' + G\,(2G_S'-1) \\
&\quad +2\sqrt{(G-1)G\,(G_S'-1)\,G_S'} + 1\Big) - 1 \\
\nu_P &= \left(\sqrt{G\kappa\,(G'-1)} - \sqrt{(G-1)\kappa G'}\right)^2 + \left(\sqrt{\eta G_S\,(G'-1)} + \sqrt{\eta\,(G_S-1)\,G'}\right)^2 - (\eta+\kappa-1)\,(G'-1) \\
c_1 &= \left(\sqrt{G\kappa\,(G'-1)} - \sqrt{(G-1)\kappa G'}\right)\left(\sqrt{\eta(G-1)\,(G_S'-1)} + \sqrt{\eta GG_S'}\right) + \\
&\quad \left(\sqrt{\eta G_S\,(G'-1)} + \sqrt{\eta\,(G_S-1)\,G'}\right)\left(\sqrt{(G_S-1)\,\kappa_S\,(G_S'-1)} - \sqrt{G_S\kappa_SG_S'}\right) - \sqrt{\eta\kappa\,(G'-1)\,G_S'} + \sqrt{\eta\kappa_S\,(G'-1)\,G_S'} \\
c_2 &= \sqrt{(G'-1)\,(G_S'-1)}\,\left(\sqrt{\eta\kappa_S} - \sqrt{\eta\kappa}\right) + \left(\sqrt{G\kappa\,(G'-1)} - \sqrt{(G-1)\kappa G'}\right)\left(\sqrt{\eta G\,(G_S'-1)} + \sqrt{\eta(G-1)G_S'}\right) \\
&\quad - \left(\sqrt{\eta G_S\,(G'-1)} + \sqrt{\eta\,(G_S-1)\,G'}\right)\left(\sqrt{G_S\kappa_S\,(G_S'-1)} - \sqrt{(G_S-1)\,\kappa_SG_S'}\right) \\
d_S &= \eta\left(2\sqrt{(G_S'-1)\,G_S'}G + 2\sqrt{(G-1)G}G_S' - \sqrt{(G-1)G} - 2\sqrt{(G_S'-1)\,G_S'}\right) + \\
&\quad \kappa_S\left(2\sqrt{(G_S'-1)\,G_S'}G_S - 2\sqrt{(G_S-1)\,G_S}G_S' + \sqrt{(G_S-1)\,G_S} - 2\sqrt{(G_S'-1)\,G_S'}\right) + \sqrt{(G_S'-1)\,G_S'} \\
d_P &= \sqrt{(G'-1)\,G'}(\eta+\kappa-1) + \left(\sqrt{G\kappa\,(G'-1)} - \sqrt{(G-1)\kappa G'}\right)\left(\sqrt{(G-1)\kappa\,(G'-1)} - \sqrt{G\kappa G'}\right) - \\
&\quad \left(\sqrt{\eta G_S\,(G'-1)} + \sqrt{\eta\,(G_S-1)\,G'}\right)\left(\sqrt{\eta\,(G_S-1)\,(G'-1)} + \sqrt{\eta G_SG'}\right)\,.
\end{aligned} \tag{D2}
$$

In the case of symmetric and lossless coupling, we can produce pure TMSV output by anti-squeezing:

$$G'^{\star} = G_S'^{\star} = \frac{1}{2}\left(\frac{2\eta(G-1)+2(G-1)\kappa+1}{\sqrt{-4\eta^2(G-1)+4\eta(G-1)((4G-2)\kappa+1)-4(G-1)\kappa^2+4(G-1)\kappa+1}}+1\right) \tag{D3}$$

for small $\eta$. Then the output covariance matrix reduces to:

$$V = \begin{pmatrix} 2\nu+1 & 0 & 2c_{\mathrm{PS}} & 0 \\ 0 & 2\nu+1 & 0 & -2c_{\mathrm{PS}} \\ 2c_{\mathrm{PS}} & 0 & 2\nu+1 & 0 \\ 0 & -2c_{\mathrm{PS}} & 0 & 2\nu+1 \end{pmatrix}, \tag{D4}$$

where

$$\nu = \frac{1}{2}\left(\sqrt{-4\eta^2(G-1)+4\eta(G-1)((4G-2)\kappa+1)+4\kappa(G(1-\kappa)+\kappa-1)+1}-1\right), \quad c_{\mathrm{PS}} = -2\sqrt{\eta(G-1)G\kappa}\,. \tag{D5}$$

## APPENDIX E: EFFECT OF ANCILLA LOSS

In this section, we explore the effect of the ancilla loss on the entanglement generation rate of our EA protocol.

In Fig. 8, we repeat the evaluation in Fig. 5 in the main text for the dual-band cooperative EA protocol, with ancilla loss increasing from red to black at step of 0.1 dB. The red curve at the top with 0 dB ancilla loss agrees with the red curve in Fig. 5. We observe a decay of performance as the ancilla loss increases to 1 dB (black), but the decay trend is continuous and moderate without any sharp drop or sudden death.

In Fig. 9, we repeat the evaluation in Fig. 6 in the main text for the dual-band cooperative EA protocol. Similarly, the red curve at the top with 0 dB ancilla loss agrees with the red curve in Fig. 6. Again, we observe a decay of performance as the ancilla loss increases to 1 dB (black), but the decay trend is continuous and moderate, considering the performances at the optimized $G$ values. The shapes of the large ancilla loss cases converge to that of the ancilla-free case in Fig. 6, since with ancilla loss approaching the coupling loss, the loss channels of probe and ancilla become symmetric and commutable with a beamsplitter, in this case, the two-mode squeezing input is equivalent to single-mode squeezing input. Here our information processing at the output is limited to anti-squeezing which does not include beamsplitter, thus the performance appears to be worse than the ancilla-free case. This implies that allowing general Gaussian information processing at the output may further improve the performance of our protocols.

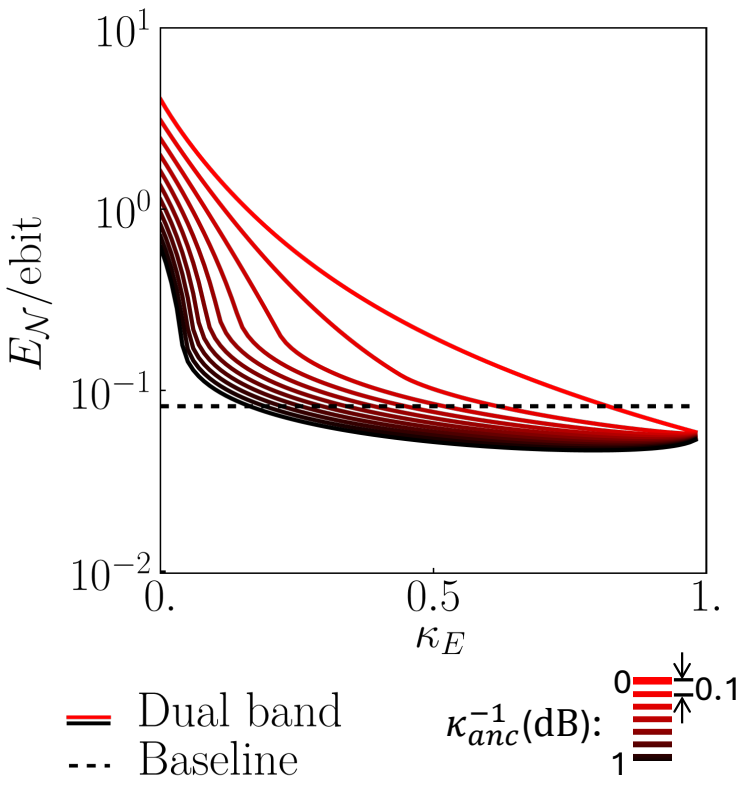

**Fig. 8.** Effect of ancilla loss on the crossband $SA$ entanglement logarithmic negativity $E_{\mathcal{N}}$ of the dual-band cooperative EA protocol versus intrinsic coupling loss $\kappa_E$, under the same setup of Fig. 5: $G$ optimized ($G_S = G$), $\eta = 1$ %. From red to black: Ancilla loss increasing from 0 dB to 1 dB at step of 0.1 dB. Ancilla loss $x$ (dB) is defined by $x = 10\log_{10}\kappa_{anc}^{-1}$, where we assume uniform ancilla loss for the two frequency bands $\kappa_A = \kappa_B = \kappa_{anc}$.

Overall, with a small ancilla loss at the level of ~0.1 dB, the advantage of our dual-band EA protocol over the baseline benchmark is still significant, especially for not too strong coupling loss $\kappa_E < 0.5$. Such small ancilla loss is not fundamentally hard, as high-fidelity quantum memory has been demonstrated with milliseconds of lifetime for continuous-variable states [54], and hours of lifetime for simple qubits [55].

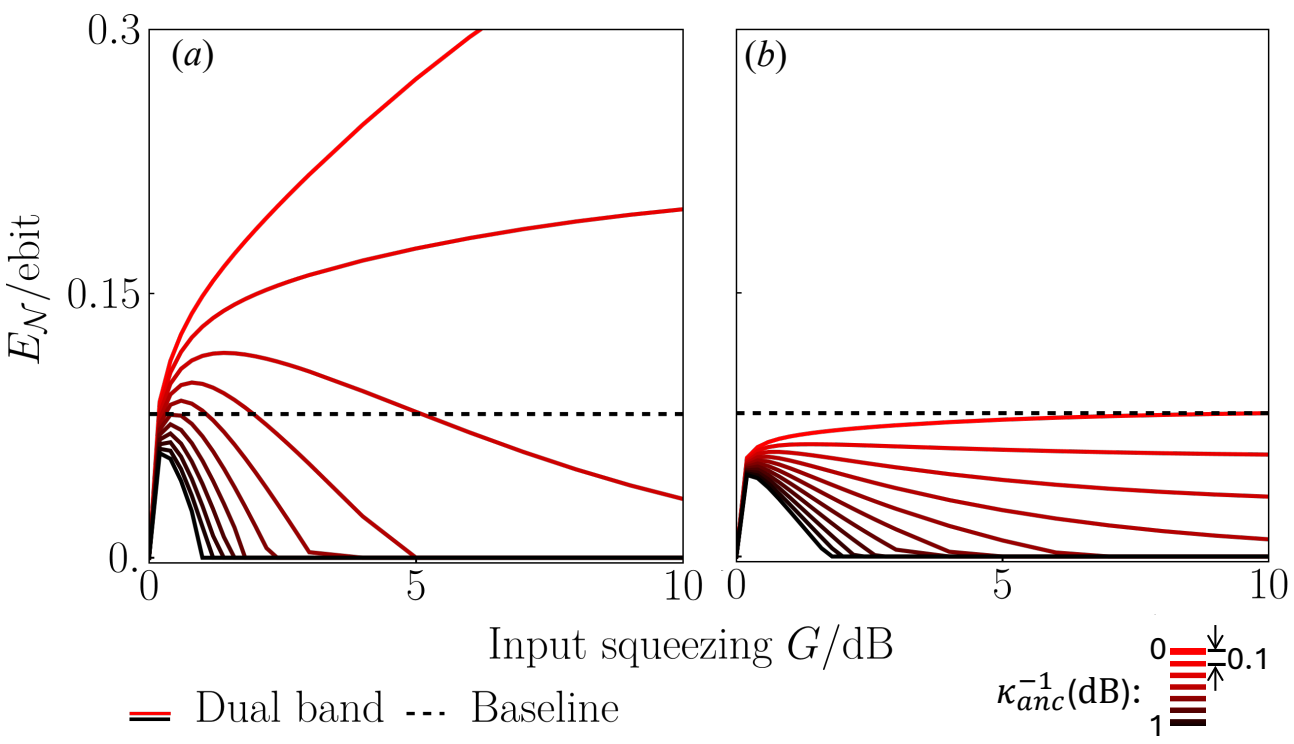

**Fig. 9.** Effect of ancilla loss on the crossband $SA$ entanglement logarithmic negativity $E_{\mathcal{N}}$ of the dual-band cooperative EA protocol versus input two-mode squeezing $G$ ($G_S = G$), under the same setup of Fig. 6: (a) $\kappa_E = 34$ %; (b) $\kappa_E = 82$ %. $\eta = 1$ %. From red to black: Ancilla loss increasing from 0 dB to 1 dB at step of 0.1 dB. Ancilla loss $x$ (dB) is defined by $x = 10\log_{10}\kappa_{anc}^{-1}$, where we assume uniform ancilla loss for the two frequency bands $\kappa_A = \kappa_B = \kappa_{anc}$.

**Funding.** U.S. National Science Foundation (CCF-2240641, OMA-2326746, 2350153); Office of Naval Research (N00014-23-1-2296); United States Air Force Office of Scientific Research (MURI FA9550-24-1-0349); Defense Advanced Research Projects Agency (HR0011-24-9-0362, HR00112490453, D24AC00153-02); Google (United States).

**Disclosures.** The authors declare no conflicts of interest.

**Data availability.** The data supporting the findings of this study are available from the first author upon reasonable request.